\documentclass[10pt]{IEEEtran}
\ifCLASSINFOpdf
\else
\fi
\usepackage{multicol}
\usepackage{enumitem,kantlipsum}
\usepackage{array}
\usepackage{amsmath}
\usepackage{graphicx,wrapfig,lipsum}
\usepackage{tabularx}
\usepackage{rotating}
\usepackage{amsmath, amssymb, amsthm}
\usepackage{stmaryrd}
\usepackage{tikz}
\usepackage{verbatim}
\usepackage{dsfont}
\usepackage{url}
\usepackage[utf8]{inputenc}
\usepackage[english]{babel}
\newtheorem{theorem}{Theorem}[]
\newtheorem{corollary}{Corollary}[]

\newtheorem{lemma}[]{Lemma}
\usepackage{multicol}
\usepackage{xr-hyper}

\usepackage[nolist,nohyperlinks]{acronym}
\usepackage[ruled,vlined]{algorithm2e}
\usepackage{algpseudocode}
\usepackage{scalerel}
\usepackage{multicol}
\usepackage[hidelinks]{hyperref}
\usepackage{mathtools}
\usepackage{authblk}
\usepackage{setspace}
\newtheorem{definition}{Definition}
\usepackage[noadjust]{cite}
\usepackage{bbm}
\usepackage[normalem]{ulem}
\usepackage{color}
\usepackage{dsfont}
\usepackage{setspace}
\usepackage{lipsum}
\usepackage{cuted}
\usepackage{amsmath}

\usepackage{subfig}
\usepackage{graphicx}

\usetikzlibrary{automata}
\usetikzlibrary{arrows}
\allowdisplaybreaks
\usepackage{algcompatible,lipsum}
\usepackage{bm}
\usetikzlibrary{automata}
\usetikzlibrary{arrows}
\allowdisplaybreaks
\usepackage{tikz}
\usepackage{acronym}
\usetikzlibrary{automata, positioning}

\begin{acronym}
    \acro{BPP}{binomial point process}
    \acro{BLP}{binomial line process}
	\acro{CDF}{cumulative density function}
    \acro{PDF}{probability density function}
    \acro{PPP}{Poisson point process}
    \acro{BS}{base station}
    \acro{LOS}{line of sight}
    \acro{SINR}{signal to interference plus noise ratio}
    \acro{SNR}{signal to noise ratio}
    \acro{mm-wave}{millimeter wave}
    \acro{PV}{Poisson-Voronoi}
    \acro{LOS}{line-of-sight}
    \acro{NLOS}{non line-of-sight}
    \acro{PLCP}{Poisson line Cox process}
    \acro{MBS}{macro base station}
    \acro{SBS}{small cell base station}
    \acro{RAT}{radio access technique}
    \acro{5G}{fifth generation}
    \acro{PLP}{Poisson line process}
    \acro{UE}{user equipment}
\end{acronym}

\title{Binomial Line Processes: Distance Distributions}

\author{Gourab Ghatak,
{\it Member, IEEE}
\vspace{-1cm}
\thanks{G. Ghatak is with the Department of ECE at IIIT Delhi. Email: gourab.ghatak@iiitd.ac.in.}}

\begin{document}
\maketitle

\begin{abstract}
    We introduce the binomial line process (BLP), a novel spatial stochastic model for the characterization of streets in the statistical evaluation of wireless and vehicular networks. Existing stochastic geometry models for streets, e.g., Poisson line processes (PLP) and Manhattan line processes (MLP) lack an important aspect of city-wide street networks: streets are denser in the city center and sparse near the suburbs. Contrary to these models, the BLP restricts the generating points of the streets to a fixed radius centered at the origin of the Euclidian plane, thereby capturing the inhomogeneity of the streets with respect to the distance from the center. We derive a closed-form expression for the contact distribution of the BLP from a random location on the plane. Leveraging this, we introduce the novel Binomial line Cox process (BLCP) to emulate points on individual lines of the BLP and derive the distribution to the nearest BLCP point from an arbitrary location. Using numerical results, we highlight that the spatial configuration of the streets is remarkably distinct from the perspective of a city center user to that of a suburban user. The framework developed in this paper can be integrated with the existing models of line processes for more accurate characterization of streets in urban and suburban environments.
\end{abstract}
\vspace{-0.7cm}
\section{Introduction}
An accurate evaluation of modern cellular and vehicular communication networks requires tractable models for the statistical characterization of urban infrastructure, e.g., streets, buildings, and blockages~\cite{maksymyuk2015stochastic, farooq2015stochastic, andrews2016primer}. This is all the more necessary when studying high-frequency \ac{mm-wave} transmissions or vehicular networks, where network properties like access point deployment, propagation, and the location of receivers are governed by the physical infrastructure~\cite{farooq2015stochastic, di2015stochastic}. In this regard, stochastic geometry based studies consider line processes for modeling streets of an urban environment~\cite{baccelli1997stochastic,6260478}. Out of the candidate line processes, the most popular one for modeling streets in wireless communication studies is the \ac{PLP}~\cite{chetlur2020stochastic, jeyaraj2020cox}. Although the \ac{PLP} model provides significant tractability in developing performance metrics of wireless networks, it fails to accurately take into account some salient features of urban street networks such as finite street lengths and inhomogeneous density of streets across a given city. The recent work by Jeyaraj {\it et al.}~\cite{jeyaraj2020cox} studied a generalized framework for Cox models to study vehicular networks. The authors significantly improved the accuracy of line process models to account for finite street lengths by considering t-junctions, stick processes, and Poisson lilypond models. However, their work does not consider inhomogeneous road densities e.g., for urban and suburban locations. In this work, we introduce a novel line process model called the \ac{BLP} which bridges this gap. It is envisaged that an accurate characterization of a city-wide street system will integrate the theory developed in this paper with the existing works~\cite{jeyaraj2020cox}. 

\subsection{Related Work on Line Processes}
Gloaguen \textit{et al.}~\cite{gloaguen2010analysis} have modeled streets using either a Poisson-line tessellation (PLT), Poisson-Voronoi tessellation, or a Poisson-Delaunay tessellation. Due to the higher tractability, the PLT (the tessellation created by the \ac{PLP}) has received significant interest in further works, as compared to the other models. For example, the \ac{PLP} was used by Morlot~\cite{6260478} to model the location of \acp{UE} and by Choi and Baccelli~\cite{choi2017analytical} to model vehicular base-stations and \acp{UE}. The \ac{PLP} was used to study on-road deployment of \ac{mm-wave} small cells in~\cite{ghatak2017modeling} and \cite{ghatak2019small}. Further information in this direction can be found in~\cite{chetlur2020stochastic}. To alleviate some of the drawbacks of a \ac{PLP}, in particular, to take into account finite length streets, the recent work by Jeyaraj {\it et al.}~\cite{jeyaraj2020cox} have provided a generalized framework for the class of line processes. They have considered the Poisson stick process and the lilypond model to introduce additional accurate structures to the street networks. Then, leveraging the same, the authors have analyzed the reliability of communication networks for these different street models.

\vspace{-0.4cm}
\subsection{Motivation: Why One More Line Process?}
\begin{figure}
    \centering
    \includegraphics[width = 0.75\linewidth]{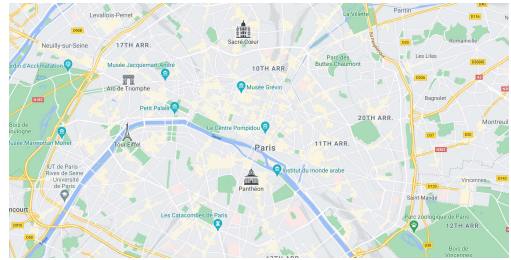}
    \caption{Part of Paris city map. Visual inspection reveals the inhomogeneity of street density.}
    \label{fig:Paris}
    \vspace{-0.2cm}
\end{figure}
Although the literature on line processes is rich, all the above models fail to take into account a particular property of the urban infrastructure: streets are denser near the city center or downtown, while they are sparse near the suburbs. This can be observed by simple visual inspection in Fig.~\ref{fig:Paris} where we depict a snapshot of Paris downtown and suburb. It is also important to note that in large cities, suburban connectivity is facilitated by longer streets that originate near the city centre. Such streets run through a number of residential clusters and often connect two cities.

\begin{figure}
\centering
\subfloat[]
{\includegraphics[width=0.3\linewidth, height = 4cm]{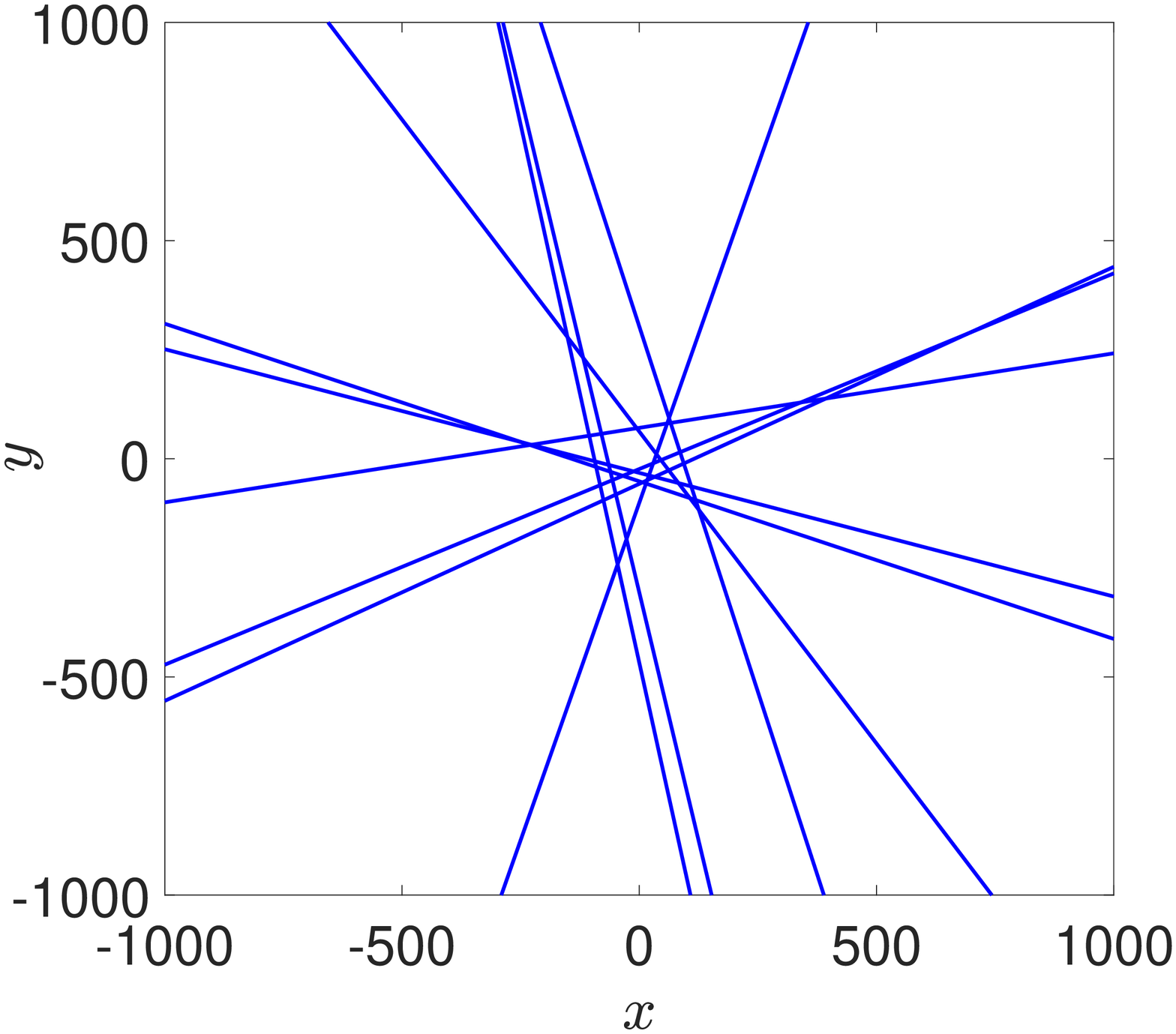}
\label{fig:blp}}
\hfil
\subfloat[]
{\includegraphics[width=0.3\linewidth, height = 4cm]{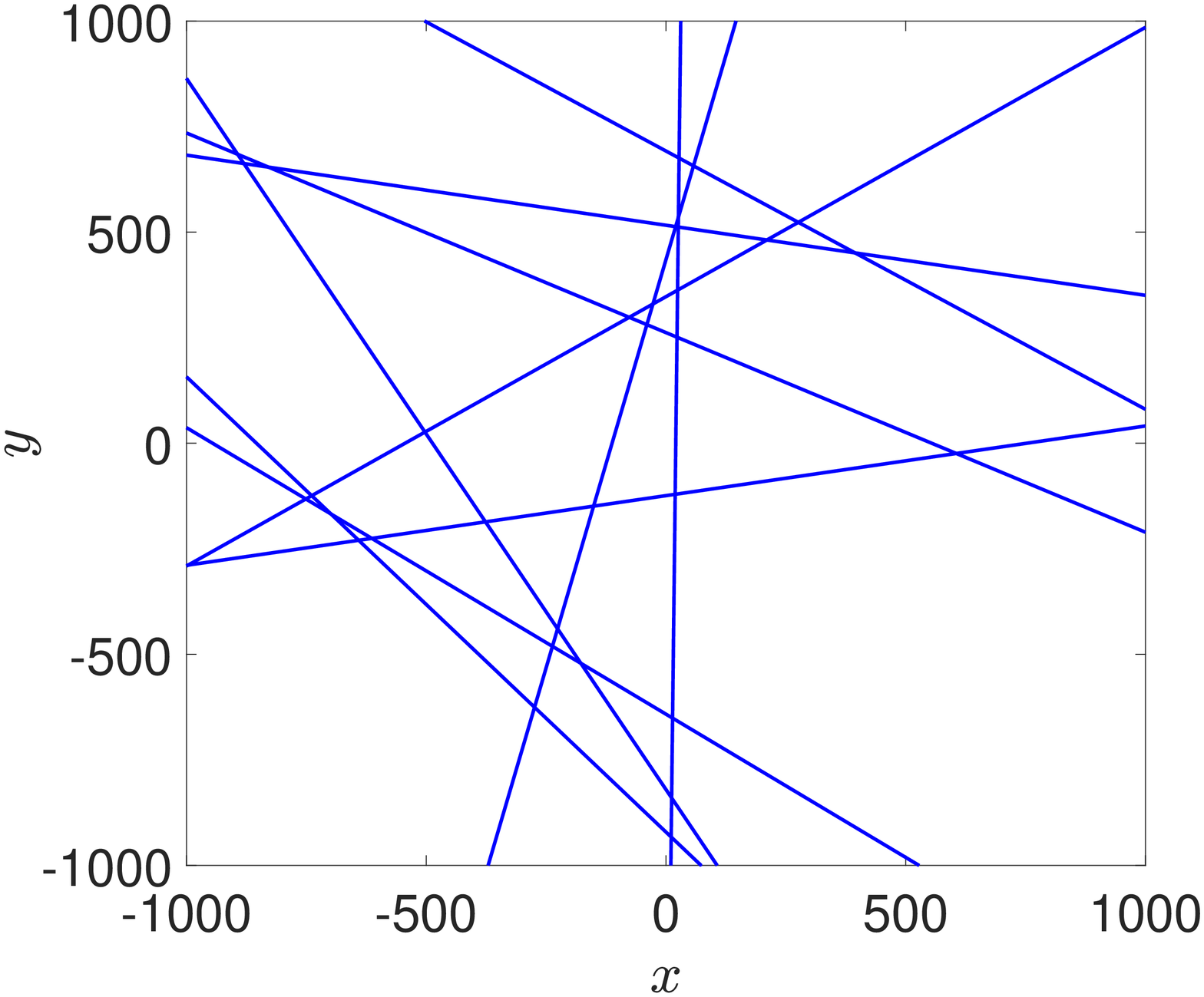}
\label{fig:plp}}
\subfloat[]
{\includegraphics[width=0.3\linewidth, height = 4cm]{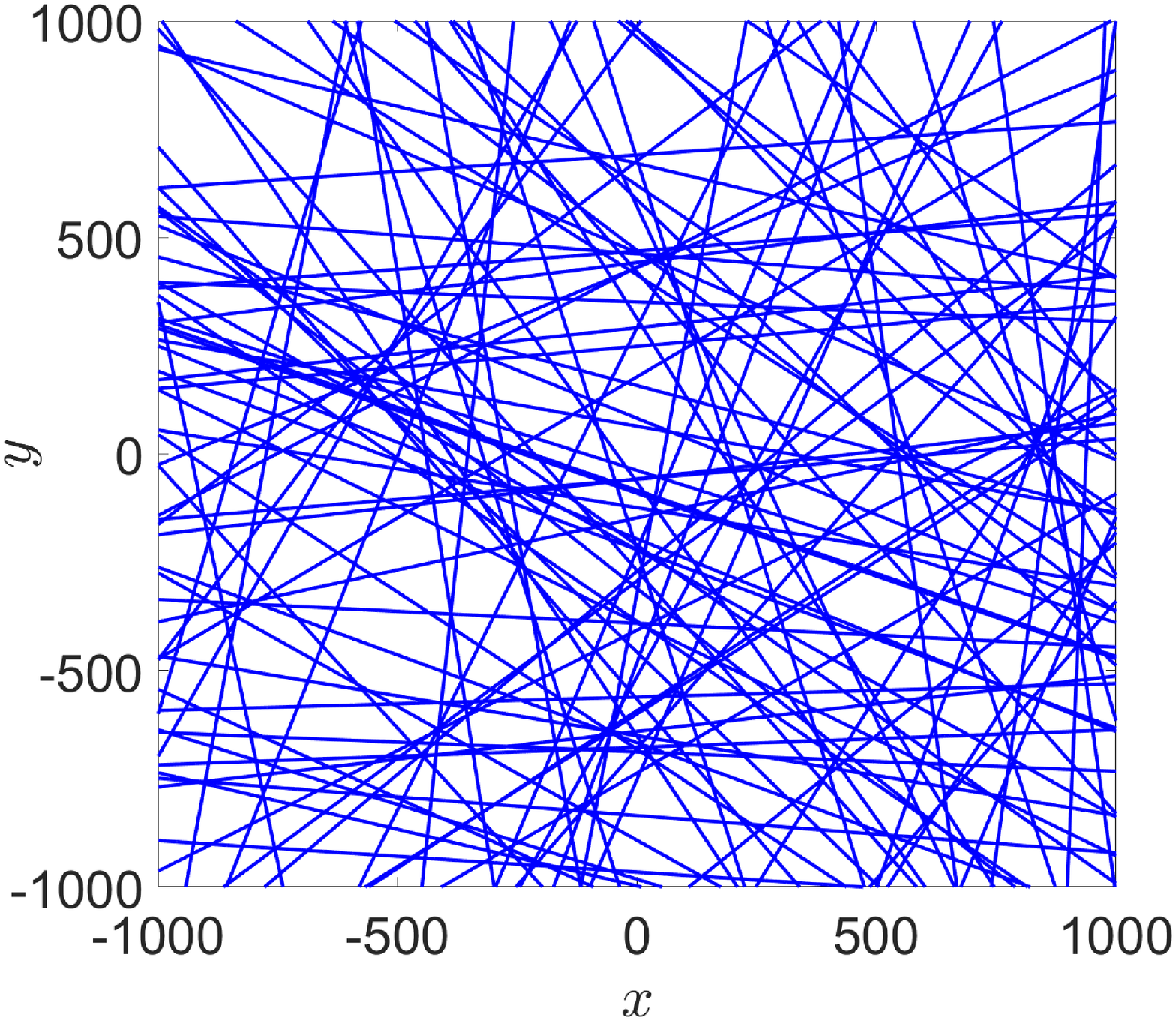}
\label{fig:plp2}}
\caption{(a) A BLP with $n_B = 10$ and $R = 100$, (b) A realization of a PLP which has 10 lines (c) A PLP with the same intensity as the BLP with $n_B = 10$.}
\label{fig:lines} 
\vspace{-0.2cm}
\end{figure}

Let us elaborate this notion with the help of realizations of two candidate line processes: PLP and BLP (Please refer to Section~\ref{sec:Cons} for details on the construction of a \ac{BLP}). In Fig.~\ref{fig:lines}, we plot a realization of a BLP (Fig.~\ref{fig:blp}), a PLP realization which has the same number of streets as that of the BLP (Fig.~\ref{fig:plp}), and a realization of a PLP which has the same street density as the BLP (Fig.~\ref{fig:plp2}). We notice that the BLP more accurately captures the two characteristics that we intend to address: i) denser roads in downtown and sparser roads in suburbs, and ii) suburbs are connected by roads that originate at the city center. Consequently, in wireless network modeling, statistical evaluation of metrics such as the distribution of the distance to the nearest street, the distance to the nearest on-street deployed access point, the number of street within a given range of a user, etc. are different for a sub-urban users and a downtown user. Thus, a study of BLP is necessary to take these nuances into account.

\begin{figure*}
\begin{align}
    x^* = \frac{1}{1 + \cot^2(\theta)} \left[r_i \cot(\theta)\csc(\theta) + x_t \pm  
     \left(-r_i^2 \csc^2(\theta) + 2 r_i x_t \csc(\theta) \cot(\theta) - x_t^2 \cot^2(\theta) +  t^2 \cot^2(\theta)  + t^2\right)^{\frac{1}{2}}\right]
    \label{eq:abscissa}
\end{align}
\hrule
\end{figure*}

\vspace{-0.4cm}
\subsection{Contributions}
The contributions of this correspondence are as follows:
\begin{itemize}
    \item We introduce and characterize the \ac{BLP}, which is a novel spatial stochastic model to characterize streets jointly in an urban and suburban environment. To the best of our knowledge, the BLP has never been studied in wireless network modeling.
    \item We introduce the notion of {\it domain bands}, which are subsets of the generating set of the BLP that correspond to lines which intersect a given disk in $\mathbb{R}^2$. We derive exact expression for the area of the domain bands. Consequently, we derive the void probability of the domain bands in the generating domain, which correspond to the probability that no lines intersect a given given disk in $\mathbb{R}^2$. We extend this framework to derive the distance distribution of the nearest line from an arbitrary point on the Euclidean plane.
    \item Leveraging the above, we introduce a doubly stochastic process called the BLCP which models locations on individual streets as a 1D \ac{PPP}. We derive the distance distribution of the nearest BLCP point from a given location in $\mathbb{R}^2$.
\end{itemize}
It must be noted that the objective of this paper is not to propose a model that completely replaces existing stochastic geometry models for characterizing streets and on-street locations in a wireless network, e.g., PLP and PLCP, respectively. On the contrary, the proposed framework can be integrated into the existing models to augment their accuracy, by jointly taking into account the relative inhomogeneity of street configurations for an urban and sub-urban user.

\section{Binomial Line Process: Construction}
\label{sec:Cons}
{A line process $\mathcal{P} \subset \mathbb{R}^2$ is a collection of lines $\{L_1, L_2, \ldots \}$ in $\mathbb{R}^2$. Let us denote a \ac{BLP} by $\mathcal{P}_B$, which is a collection of a fixed number $n_B$ of lines. Any line that belongs to $\mathcal{P}_B$ is uniquely characterized by the distance $r_i$ between the origin $O$ and its projection $P$ on the line, and by the angle $\theta$ between $\vec{OP}$ and the x-axis on the other hand. For a \ac{BLP}, the domain of the pair of parameters $(\theta, r_i)$ is the finite half cylinder $\mathcal{D}:=$ [$0,\pi$) $\times$ $[-R, R]$. We will call $\mathcal{D}$ as the generating set or the domain set of $\mathcal{P}_B$, and a point $(\theta_i, r_i)  \in \mathcal{D}$, corresponding to a line $L_i \in \mathcal{P}_B$, the generating point of $L_i$. Accordingly, there is a bijective mapping $f:\mathcal{P}_B\rightarrow \mathcal{D}$ between any random point $(\theta_i , r_i) \in \mathcal{D}$ and a corresponding line $L_i \in \mathcal{P}_B$. We can now define a \ac{BLP}, formally.
\begin{definition}
A line process $\mathcal{P}_B\triangleq \{L_i\}$ in $\mathbb{R}^2$ consisting of $n_B$ lines is a \ac{BLP}, if and only if the set of corresponding $n_B$ generating points $\{(\theta_i , r_i)=f(L_i)\}$ is a \ac{BPP} in $\mathcal{D}$.
\end{definition}
}
\begin{figure}
    \centering
    \includegraphics[width = 0.75\linewidth]{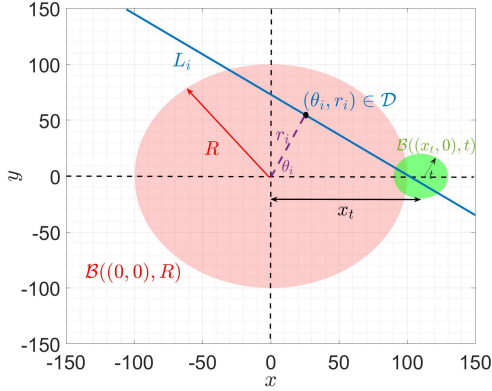}
    \caption{Illustration of the construction of a BLP and intersecting lines on $\mathcal{B}((x_t, 0), t)$.}
    \label{fig:Construction}
\end{figure}
This is elaborated further in Fig.~\ref{fig:Construction}, where the generating points of the lines are restricted to the disk $\mathcal{B}((0,0),R)$. It is important to note that unlike stationary point processes, the statistics of the \ac{BLP} cannot be studied from the perspective of a typical point. However, due to the isotropic construction of the \ac{BLP}, we can assume that the properties of the \ac{BLP} as seen from a point depends only on its distance from the center and not its orientation. Accordingly, without loss of generality, let us consider a test point located at $(x_t, 0)$. First, we study the set of lines that intersect a randomly located disk in $\mathbb{R}^2$ centered around a test point $(x_t,0)$. 

\vspace{-0.4cm}
\subsection{Domain bands in $\mathcal{D}$}
The equation of the line $L_i$ corresponding to $(\theta_i, r_i) \in \mathcal{D}$ is given by:
    $x \cos(\theta_i) + y \sin(\theta_i) = r_i$.
On the other hand, the equation of the circle $\mathcal{B}_p((x_t,0), t)$ with the test node at its center and a radius of $t$ is given by:
    $(x - x_t)^2 + y^2 = t^2$.
Now, solving the above two equations simultaneously, we obtain the abscissa of the intersection points of $L_i$ and $\mathcal{B}_p((x_t,0), t)$ as \eqref{eq:abscissa}, depicted on the top of the next page.

In order to find the set of $(\theta_i, r_i)$ for which $L_i$ intersects $\mathcal{B}((x_t,0), t)$, we need to find the $(\theta_i, r_i)$ which results in $L_i$ being a tangent to $\mathcal{B}((x_t,0), t)$. Let us note that for a given $\theta_i$, the value(s) of $r_i$ for which \eqref{eq:abscissa} has only one solution is obtained by solving:
    $-r_i^2 \csc^2(\theta) + 2 r_i x_t \csc(\theta) \cot(\theta) - x_t^2 \cot^2(\theta) 
    +  t^2 \cot^2(\theta)  + t^2 = 0$,
which results in:
\begin{align}
    r_i^*(\theta) = x_t \cos(\theta) \pm t
    \label{eq:domain}
\end{align}
Let the two solutions above be represented by $r_1(\theta)$ and $r_2(\theta)$ for the positive and the negative signs, respectively. In other words, all the points $(\theta_i , r_i) \in \mathcal{D}$ that fall within the set given by \eqref{eq:domain} generate lines $L_i$ that intersect $\mathcal{B}((x_t,0), t)$.
\begin{figure}
    \centering
    \includegraphics[width = 0.75\linewidth]{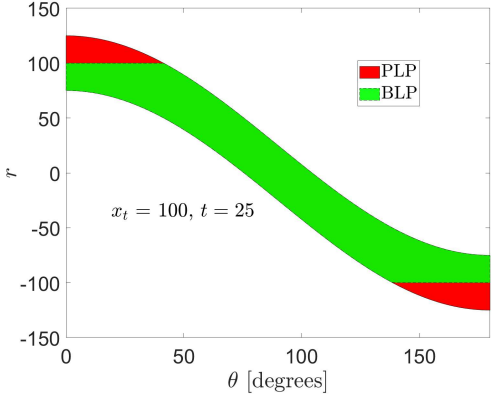}
    \caption{Set of points in $\mathcal{D}$ which generate lines in $\mathbb{R}^2$ that intersect $\mathcal{B}((x_t,0), t)$ for a BLP (denoted by green) and a PLP (denoted by red). Note that the part of the red region behind the green region is hidden. Here $R = 100$.}
    \label{fig:Domain0}
    \vspace{-0.3cm}
\end{figure}
In Fig.~\ref{fig:Domain0}, we show the set of points in $\mathcal{D}$ that generate lines in $\mathbb{R}^2$ intersecting $\mathcal{B}_p((x_t,0), t)$. Due to their structure, henceforth they are referred to as {\it domain bands}. It is interesting to note that the domain band for the BLP is a clipped version of the PLP due to the restriction of the points to lie within $\mathcal{B}((0,0), R)$.
\begin{figure}
    \centering
    \includegraphics[width = 0.75\linewidth]{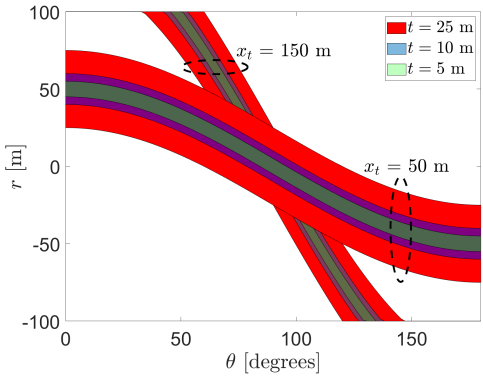}
    \caption{Domain bands for different values of $t$ and $x_t$. Here $R = 100$. Note that when $|x_t| + t \leq R$, the domain bands for PLP and BLP coincide.}
    \label{fig:Domain1}
    \vspace{-0.3cm}
\end{figure}
In Fig.~\ref{fig:Domain1} we plot the domain bands for the BLP for different values of $x_t$ and $t$. We note that when $|x_t| + t < R$, the domain bands for \ac{BLP} and \ac{PLP} coincide. Additionally, the width of the band decreases as $t$ decreases or $x_t$ increases. As we will see in the next section.
\vspace{-0.4cm}
\subsection{Void Probabilities}
\begin{corollary}
The area of the domain band for a PLP is independent of $x_t$.
\end{corollary}
This can be easily verified by evaluating the following:
\begin{align}
    \int_0^\pi \sin^2(\theta) \left[x_t \cot(\theta) \csc(\theta)\right] + t \csc^2(\theta) {\rm d}\theta - \nonumber \\
    \int_0^\pi \sin^2(\theta) \left[x_t \cot(\theta) \csc(\theta)\right] - t \csc^2(\theta) {\rm d}\theta = 2 \pi t.
\end{align}
The classical result of PLP follows from above:
\begin{corollary}
The number of lines of a PLP intersecting $\mathcal{B}((0,x_t),t)$ is Poisson distributed with parameter $2\pi t$. Accordingly, the probability that no lines intersect $\mathcal{B}((0,x_t),t)$ is given by $\exp(- 2 \pi \lambda t)$.
\end{corollary}

For a BLP, however, the area of the domain band needs to be calculated explicitly. For the same, let us first note that the values of $\theta_i$ for which \eqref{eq:domain} exceeds $\pm R$ is given by:
\begin{align}
    \theta_{11}(x_t, t) &= \cos^{-1}\left(\frac{R - t}{x_t}\right), 
    &\theta_{12}(x_t, t) = \cos^{-1}\left(-\frac{R + t}{x_t}\right)\nonumber \\
    \theta_{21}(x_t, t) &=  \cos^{-1}\left(\frac{R + t}{x_t}\right),
    &\theta_{22}(x_t, t) = \cos^{-1}\left(-\frac{R - t}{x_t}\right)\nonumber
\end{align}

We have now developed the necessary tools to evaluate the area of the domain band for a BLP, which is presented below.
\begin{theorem}
The area of the domain band for a BLP $\mathcal{P}_B$ defined on $[0,\pi)\times [-R,R]$ corresponding to $\mathcal{B}(x_t, t)$ is:

\begin{align}
    &\mathcal{A}_D(x_t,t) = \nonumber \\
    &\begin{cases}
        2 \pi t ; \quad \nonumber \text{for } |x_t| + t \leq R \\
        2 \left[\pi t - x_t \sqrt{1 - \left(\frac{R - t}{x_t}\right)^{2}} +
        \left(R - t\right)  \cos^{-1} \left(\frac{R - t}{x_t}\right)\right]; \nonumber \\
        \hspace{1cm} \text{for } |x_t| + t > R \text{ and } |x_t| - t \leq R 
        \nonumber \\
        2 \left[\pi t - x_t \left(\sqrt{1 - \left(\frac{R - t}{x_t}\right)^{2}} 
         - \sqrt{1 - \left(\frac{R + t}{x_t}\right)^{2}} \right) \right.\\
         \hspace{0.4 cm} \left. + \left(R - t\right)  
         \cos^{-1} \left(\frac{R - t}{x_t}\right)
        - \left(R + t\right)  \cos^{-1} \left(\frac{R + t}{x_t}\right)\right];\\
        \hspace{1cm} \text{for } |x_t| - t \geq R \nonumber
    \end{cases}
\end{align}
\end{theorem}

\begin{IEEEproof}
\begin{figure}
\centering
\subfloat[]
{\includegraphics[width=0.3\linewidth, height = 4cm]{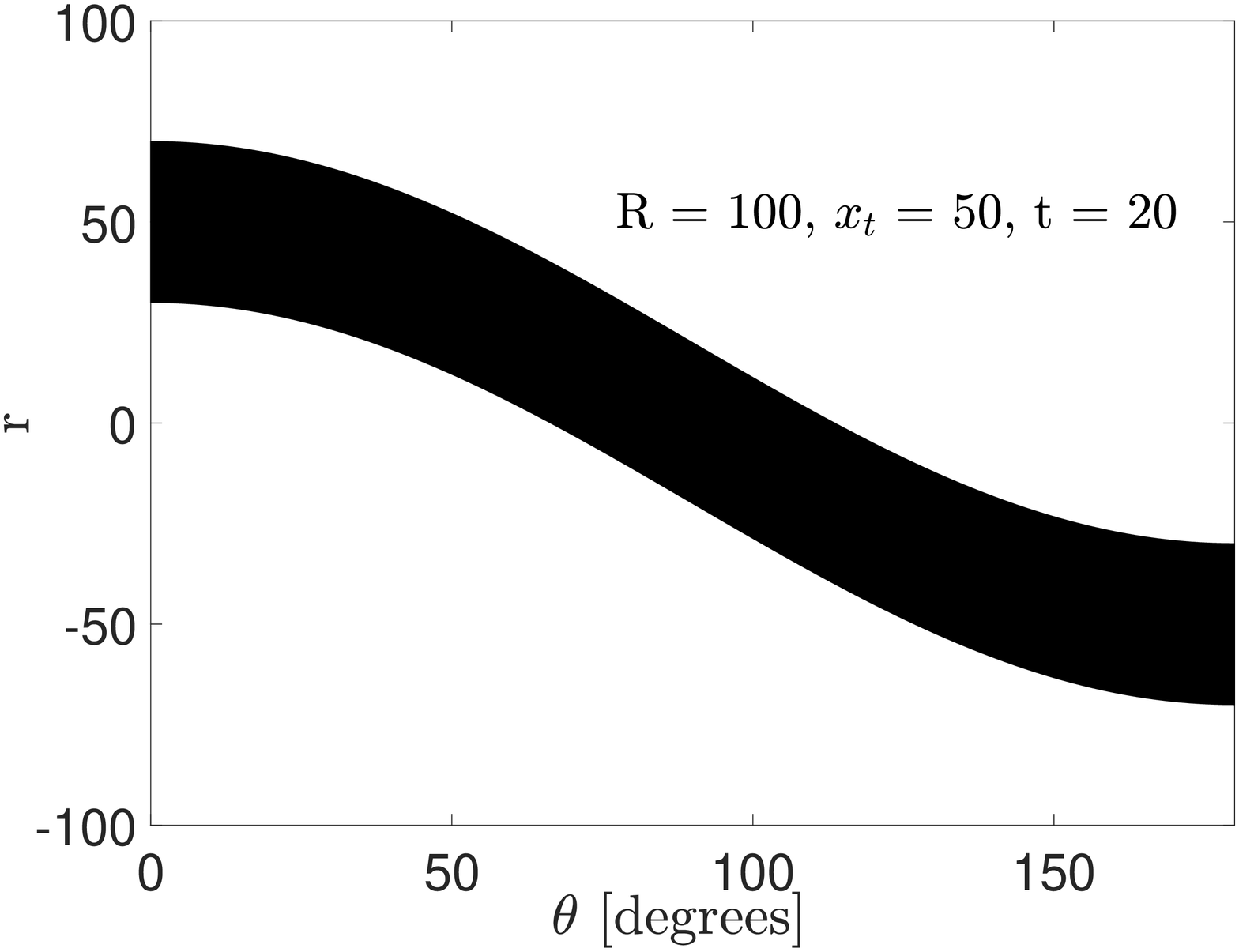}
\label{fig:Case1}}
\hfil
\subfloat[]
{\includegraphics[width=0.3\linewidth, height = 4cm]{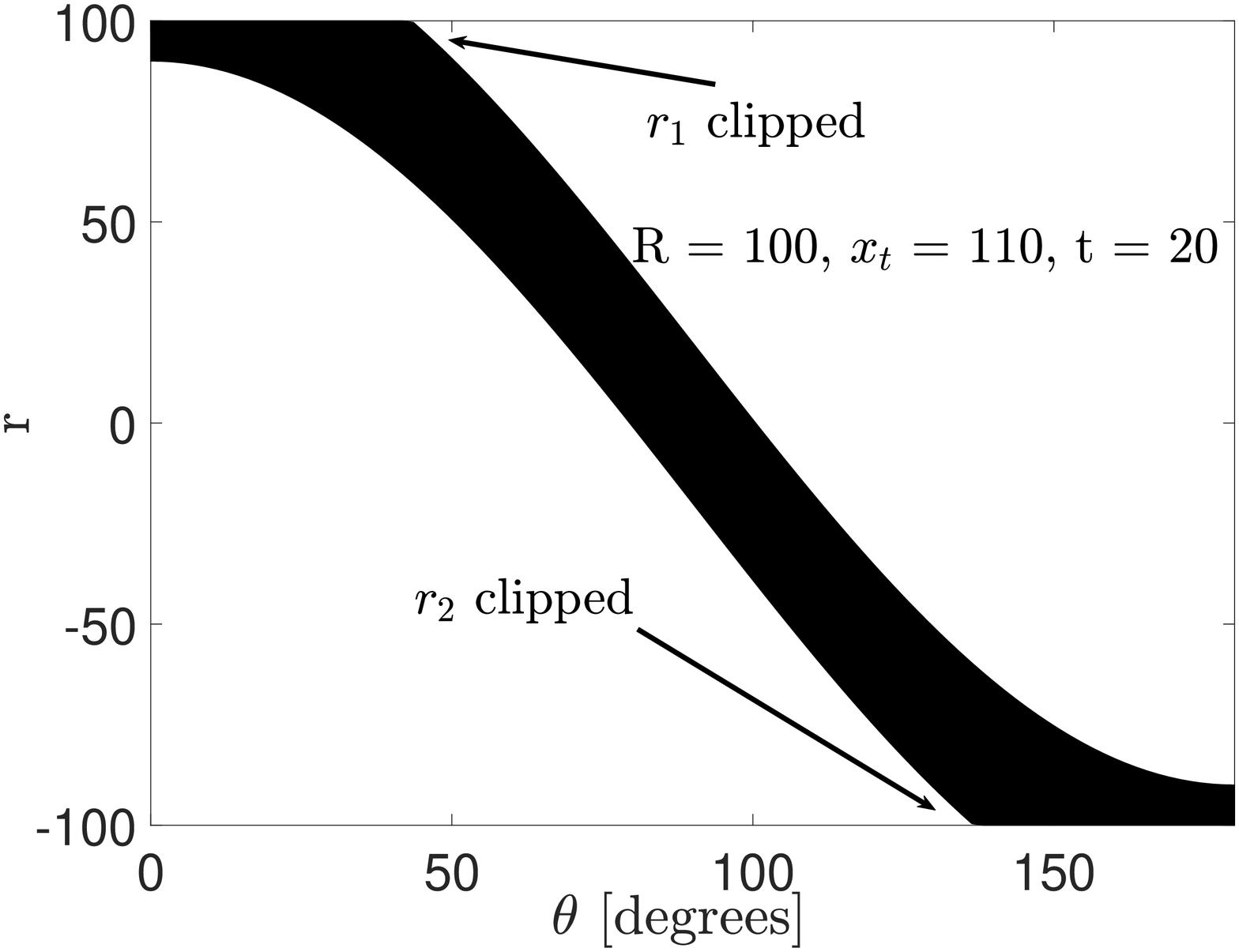}
\label{fig:Case2}}
\subfloat[]
{\includegraphics[width=0.3\linewidth, height = 4cm]{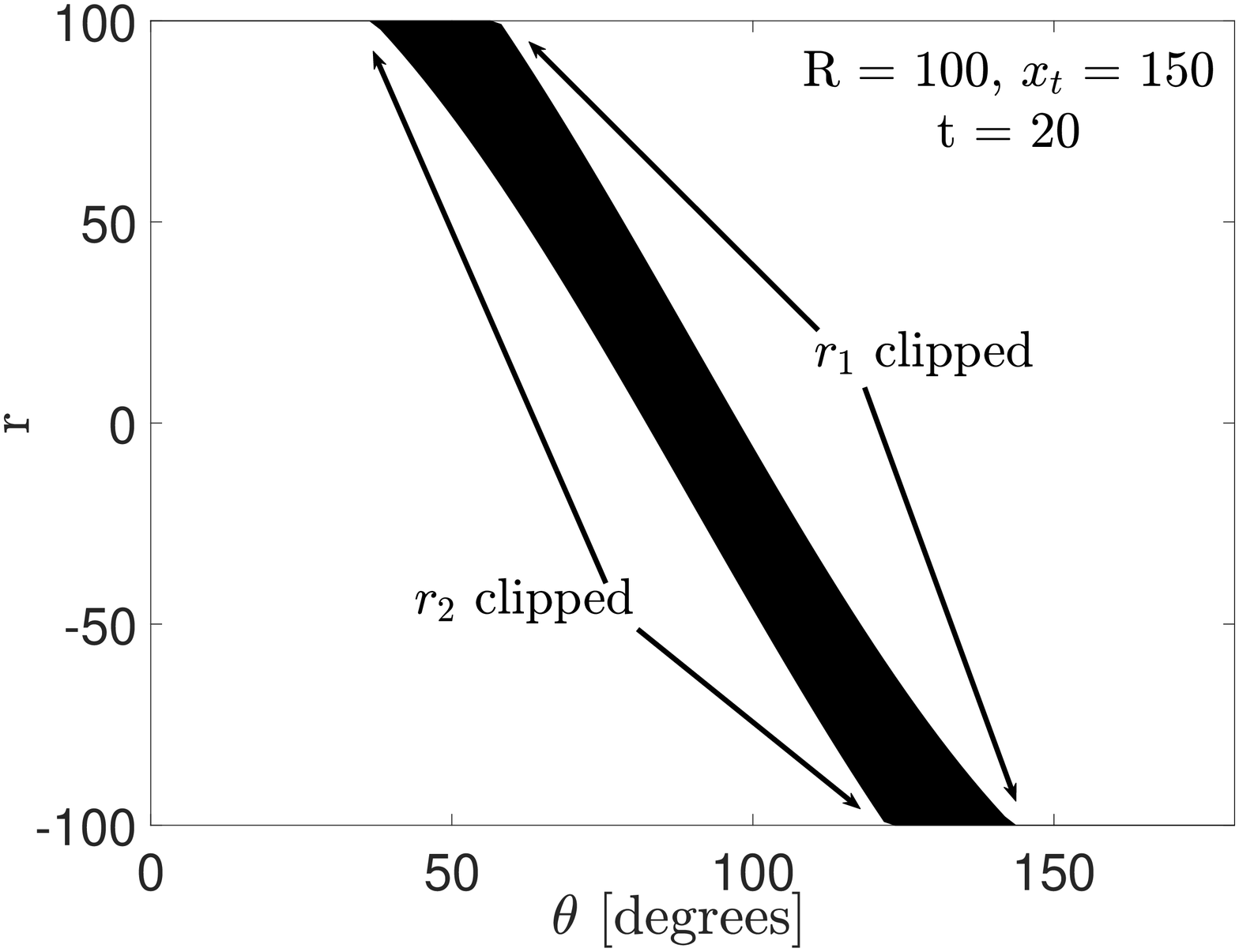}
\label{fig:Case3}}
\caption{(a) Case 1: $|x_t| + t \leq R$ (b) Case 2 $|x_t| + t \geq R$ and $|x_t| - t \leq R$ and (c) Case 3: $|x_t| - t \geq R$.}
\label{fig:Cases} 
\end{figure}
Let us consider three cases below:

{\bf Case 1: $|x_t| + t \leq R$.} (Refer to Fig.~\ref{fig:Case1}). Here, naturally we have $|x_t| - t \leq R$. For this case, the domain band for the BLP coincides with that of the PLP. Accordingly, the area of the domain band for Case 1 is $\mathcal{A}_{D1}(x_t,t) =  2 \pi t$.

{\bf Case 2:} $|x_t| + t > R$ and $|x_t| - t \leq R$.(Refer to Fig.~\ref{fig:Case2}). In this case, the upper part of the band is clipped on the left, and the lower part of the band is clipped on the right. Noting the symmetry of the domain bands from the perspective of the $\theta = \frac{\pi}{2}$ line, the area is evaluated as:
\begin{align}
    \mathcal {A}_{D2}(x_t, t) &=  2 \pi t - 2\int_0^{\cos^{-1} \left(\frac{R - t}{x_t}\right)} 
         \left(x_t \cos(\theta) + t \right) {\rm d}\theta \nonumber\\
     &+ 2 \int_0^{\cos^{-1} \left(\frac{R - t}{x_t}\right)} R {\rm d} \theta \nonumber
\end{align}

{\bf Case 3: $|x_t| - t > R$. }(Refer to Fig.~\ref{fig:Case3}). Here, naturally, $|x_t| + t > R$. In this case, both the lower and the upper part of the domain band is clipped on both left and right ends, so in a similar manner to the above case, the area of the domain band is: \\

\begin{align}
     &\mathcal{A}_{D3}(x_t, t)= \nonumber \\ 
    & \mathcal \quad  2 \pi t - 2\int_0^{\cos^{-1} 
     \left(\frac{R + t}{x_t}\right)} 
      \left[\left(x_t \cos(\theta) + t \right)
         - \left(x_t \cos(\theta) - t \right) \right]
     {\rm d}\theta \nonumber\\
     &- \, 2 \int_{\cos^{-1} \left(\frac{R + t}{x_t}\right)} ^ {\cos^{-1} \left(\frac{R - t}{x_t}\right)} \left(x_t \cos(\theta) + t \right) 
     {\rm d} \theta
     + 2 \int_{\cos^{-1} \left(\frac{R + t}{x_t}\right)}^
     {\cos^{-1} \left(\frac{R - t}{x_t}\right)} R \,{\rm d} \theta \nonumber
\end{align}
Evaluating the above integrals for Case 2 and Case 3 completes the proof.
\end{IEEEproof}

\begin{corollary}
The probability that no line of the BLP intersects with $\mathcal{B}((x_t,0), t)$ is given by:
\begin{align}
    \mathcal{V}_B\left(\mathcal{B}((x_t,0), t)\right) =  \left(\frac{2 \pi  R - \mathcal{A}_\mathcal{D}\left(x_t, t\right)}{2 \pi R}\right)^{n_B}
\end{align}
\end{corollary}
Consequently, the \ac{CDF} of the distance to the nearest line from $(x_t,0)$ is given in the following lemma.
\begin{lemma}
For the point $(x_t, 0)$, the \ac{CDF} of the distance to the nearest line of the BLP is given by:
\begin{align}
    F_d(t) = 1 -  \mathcal{V}_B\left(\mathcal{B}((x_t,0), t)\right) \nonumber 
\end{align}
\end{lemma}

\begin{figure}
    \centering
    \includegraphics[width = 0.75\linewidth]{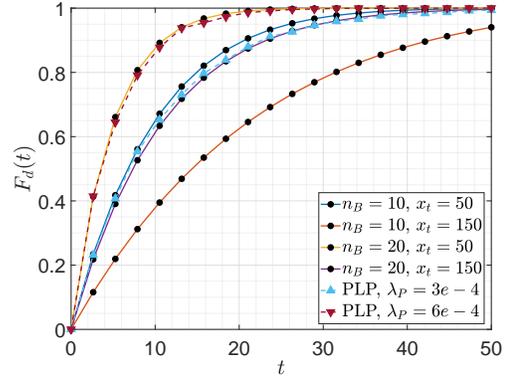}
    \caption{Distance distribution of the nearest line from $(x_t, 0)$. The solid lines represent the analytical expressions and the markers correspond to the Monte-Carlo simulations.}
    \label{fig:cdf_line_plot}
    \vspace{-0.3cm}
\end{figure}
In Fig.~\ref{fig:cdf_line_plot} we plot the CDF for different values of $n_B$ and $x_t$. For comparison, we have also included the CDF of the nearest line of a \ac{PLP} with the same intensity $\lambda^{PPP}$ as that of a \ac{BLP} ($\frac{n_B}{ 2 \pi R}$) with $n_B = 10$ and $n_B = 20$ points for $R = 100$. Noticeably the nearest line is statistically closer for a larger value of $n_B$ and lower value of $x_t$. This inference is missed by the CDF of the distance to the nearest line in a \ac{PLP}, where the CDFs coincide for all values of $x_t$ for a given $\lambda^{PPP}$.

\section{On Binomial Line Cox Processes}
For each line $L_i \in \mathcal{P}_B$, let us define a 1D \ac{PPP} $\Phi_i$ with intensity $\lambda$. We assume that $\Phi_i$ is independent of $\Phi_j$ for $L_i \neq L_j$. The set of all such points created is defined as a BLCP. Next, we derive the distance distribution to the nearest BLCP point from $(x_t, 0)$. For that, let us first characterize the length of a randomly oriented chord created by $(\theta_i, r_i) \in \mathcal{D}$ and $\mathcal{B}((x_t, 0), t)$. Let the two solutions of \eqref{eq:abscissa} be denoted by $x_1^*$ and $x_2^*$, respectively. Then, the corresponding ordinates are evaluated as $y_i^* = \frac{r_i - x_i^* \cos(\theta)}{\sin(\theta)}$, $i = \{1, 2\}$. The length of the chord thus is:
$    C_i(x_t, t, r_i, \theta_i) = ||(x^*_1,y^*_1) - (x^*_2,y^*_2)||_2. 
$
Applying the condition that $(\theta_i, r_i)$ has to lie within the BLP domain band, we denote:
$r'_1(\theta) = \max(0, \min(R , r_1(\theta)))$ and  
$r'_2(\theta) =  \max(0, \min(R , r_2(\theta))).$
The following theorem derives the distance distribution of the nearest BLCP point.
\begin{theorem}
The \ac{CDF} of the distance to the nearest BLCP point from $(x_t,0)$ is given by:
\begin{align}
   F^{BLCP}_{d}(t) = & 1 - \mathcal{P}_C(x_t, t)  \left(\frac{1}{2\pi R}\right)^{n_B - 1} \nonumber \\
   &\left[\frac{\mathcal{T}(x_t, t)^{n_B} - \left(2\pi R - \mathcal{A}_\mathcal{D}(x_t,t)\right)^{n_B}}{\mathcal{T}(x_t, t) - \left(2\pi R - \mathcal{A}_\mathcal{D}(x_t,t)\right)}\right]
\end{align}
\end{theorem}
where 
\begin{align}
\mathcal{T}(x_t, t) = \int_0^{\pi}\int_{r'_2}^{r'_1} \exp\left(-\lambda_B C_i(x_t, t, r_i, \theta_i)\right) {\rm d}r {\rm d}\theta, \nonumber 
\end{align} 
and
\begin{align}
&\mathcal{P}_C(x_t, t) = \frac{1}{\pi}\int_{\frac{\pi}{2}}^{\frac{3\pi}{2}}\left[\frac{1}{\mathcal{A}_\mathcal{D}(x_t,t)} \cdot \right. \nonumber \\
&\left.\int_0^\pi \exp \left( - \lambda C'(x_t, t,r_{\phi}, \theta_\phi)\right) {\rm d}\theta_\phi\right] {\rm d}\phi .  
\end{align}
Here, $C'(x_t, t, r_{\phi}, \theta_\phi)$ is the length of a chord which passes through the nearest point of the BLCP. It is evaluated as:
\begin{align}
    C'(x_t, t,r_{\phi}, \theta_\phi) = ||(x_0, y_0) - (x'_1,y'_1) ||_2
\end{align}
where,
\begin{align}
    x_0 & = x_t + t\cos(\theta_\phi), \quad
    y_0  = t\sin(\theta_\phi) \nonumber \\
    r_\phi & = x_0 \cos(\theta_\phi) + y_0 \sin(\theta_\phi) \nonumber \\
    x'_1 & =  \frac{1}{1 + \cot^2(\theta)} \left[r_\phi \cot(\theta_\phi)\csc(\theta_\phi) + x_t + 
     \left(-r_\phi^2 \csc^2(\theta_\phi) \right.\right. \nonumber \\
     & \left.\left. + 2 r_\phi x_t \csc(\theta_\phi) \cot(\theta_\phi) - x_t^2 \cot^2(\theta) +  t^2 \cot^2(\theta)  + t^2\right)^{\frac{1}{2}}\right] \nonumber \\
      y'_1 & =  \frac{r_\phi - x'_1 \cos(\theta_\phi)}{\sin(\theta_\phi)} \nonumber 
\end{align}
\begin{IEEEproof}
\begin{figure}
    \centering
    \includegraphics[width = 0.75\linewidth]{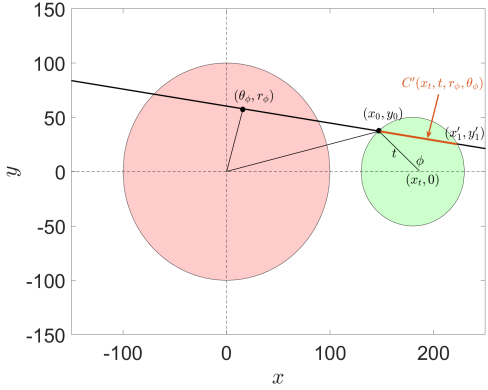}
    \caption{Illustration of the chord length $C'(x_t,t,r_\phi, \theta_\phi)$.}
    \label{fig:phi}
    \vspace{-0.3cm}
\end{figure}
Let the nearest point from $(x_t, 0)$ be at a distance $t$ from it and is at $(x_0, y_0)$. Thus, there are no BLCP points within $\mathcal{B}((x_t,0),t)$. Accordingly, we have the following two events: {\bf Event 1:} There are no points on the randomly oriented chord formed by $\mathcal{B}((x_t,0),t)$ and the line that passes through $(x_0, y_0)$. {\bf Event 2:} There are no points on the chords formed by the remaining $n_B - 1$ lines on $\mathcal{B}((x_t, 0), t)$.

Let us evaluate the probability of the first event. The point $(x_0,y_0)$ can be at a random orientation $\phi$ with respect to $(x_t,0)$, where $x_0 = x_t + t \cos(\phi)$ and $y_0 = t\sin(\phi)$. Let the other end point of the chord passing through $(x_0 , y_0)$ be denoted by $(x'_1, y'_1)$. In order to avoid double counting of the same chord, without loss of generality let us assume that $x_0 \leq x'_1$ and accordingly, $\frac{\pi}{2} \leq \phi \leq \frac{3\pi}{2}$. This is illustrated in Fig.~\ref{fig:phi}. The orientation $\phi$, and the set of corresponding points in the domain band governs the length of the chord $C'(x_t, t, r_\phi, \theta_\phi)$. The equation for such a line is $r_\phi = x_0 \cos(\theta_\phi) + y_0 \sin(\theta_\phi)$. Note here that for a given $\theta_\phi$ there exists at most one value of $r_\phi$ which results in a chord passing through $(x_0, y_0)$. Substituting this in place of $r_i$ in \eqref{eq:abscissa}, we obtain the abscissa $x'_i, i = 1,2$ of the end-points of the chord. Then the ordinates are evaluated as $y'_i = \frac{r_\phi - x'_i \cos(\theta_\phi)}{\sin(\theta_\phi)}$. Only the points $(\theta_\phi, r_\phi)$ for a given $\phi$ that belong to the domain band will form chords with positive lengths. The void probability of the BLCP on the chord $C'(x_t, t, r_\phi, \theta_\phi)$ is given by $\exp\left(- \lambda C'(x_t, t, r_\phi, \theta_\phi)\right)$. Then, de-conditioning on all possible values of $r_\phi$ and $\theta_\phi$, for all values of $\phi$, we obtain the value of $\mathcal{P}_C(x_t, t)$.

Next, apart from the tagged line, there are potentially $n_B - 1$ lines, $\{L_1, L_2, \ldots, L_{n_B - 1}\}$ that intersect $\mathcal{B}(x_t , t)$. Let $L_i$ be generated by $(r_i, \theta_i)$ and its corresponding chord be denoted by $C_i(x_t,t,r_i, \theta_i)$. Then, the probability that $C_i(x_t,t,r_i, \theta_i)$ does not contain any points from the BLCP is given as: $\exp\left(-\lambda_B C_i(x_t,t,r_i, \theta_i) \right)$. Conditioned on $n \leq n_B$ points in the domain band, the probability that no BLCP points fall in the $n_B - 1$ non-tagged lines is evaluated as:
\begin{align}
    &\sum_{n = 0}^{n_B - 1} \left(\frac{\mathcal{A}_\mathcal{D}\left(x_t, t\right)}{2 \pi R}\right)^{n} \left(1 - \frac{\mathcal{A}_\mathcal{D}\left(x_t, t\right)}{2 \pi R}\right)^{n_B - n - 1} \nonumber \\
    &\left(\frac{1}{\mathcal{A}_\mathcal{D}(x_t,t)}\int_0^{\pi}\int_{r'_2}^{r'_1} \exp\left(-\lambda_B C_i(x_t, t, r_i, \theta_i)\right) {\rm d}r {\rm d}\theta\right)^n \nonumber
\end{align}
Solving the above summation completes the proof.
\end{IEEEproof}
\begin{figure}
    \centering
    \includegraphics[width = 0.75\linewidth]{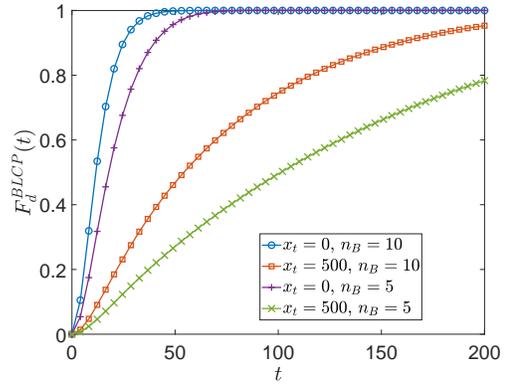}
    \caption{The distance distribution to the nearest BLCP point from $(x_t, 0)$. Here, $R = 100$ and $\lambda = 0.05$.}
    \label{fig:cdf_blcp_plot}
    \vspace{-0.3cm}
\end{figure}
Fig.~\ref{fig:cdf_blcp_plot} shows that statistically the nearest BLCP point is closer to $(x_t, 0)$ for a lower value of $x_t$ and a higher values of $n_B$. The nearest BLCP point is an important parameter while studying user association, e.g., for RSSI based base-station selection. Currently, we are investigating probability generating functional of the BLCP and its applications in modeling wireless performance metrics such as \ac{SINR} coverage probability and network throughout. This will be reported in a future work.


\vspace{-0.3cm}
\section{Discussion and Conclusions}
By deriving exact expressions for the domain bands, we have shown how the distance distributions may significantly differ for a city-center user as compared to a sub-urban user. An accurate system design, dimensioning, and initial deployment planning of wireless networks must take this nuance of streets into account which the classical models miss. The framework developed in this paper can be integrated with the existing spatial stochastic models to augment the accuracy in studying wireless and vehicular networks. 

\vspace{-0.3cm}
\bibliography{references.bib}

\begin{thebibliography}{10}
\providecommand{\url}[1]{#1}
\csname url@samestyle\endcsname
\providecommand{\newblock}{\relax}
\providecommand{\bibinfo}[2]{#2}
\providecommand{\BIBentrySTDinterwordspacing}{\spaceskip=0pt\relax}
\providecommand{\BIBentryALTinterwordstretchfactor}{4}
\providecommand{\BIBentryALTinterwordspacing}{\spaceskip=\fontdimen2\font plus
\BIBentryALTinterwordstretchfactor\fontdimen3\font minus
  \fontdimen4\font\relax}
\providecommand{\BIBforeignlanguage}[2]{{%
\expandafter\ifx\csname l@#1\endcsname\relax
\typeout{** WARNING: IEEEtran.bst: No hyphenation pattern has been}%
\typeout{** loaded for the language `#1'. Using the pattern for}%
\typeout{** the default language instead.}%
\else
\language=\csname l@#1\endcsname
\fi
#2}}
\providecommand{\BIBdecl}{\relax}
\BIBdecl

\bibitem{maksymyuk2015stochastic}
T.~Maksymyuk \emph{et~al.}, ``{Stochastic geometry models for 5G heterogeneous
  mobile networks},'' \emph{SmartCR}, vol.~5, pp. 89--101, 2015.

\bibitem{farooq2015stochastic}
M.~J. Farooq \emph{et~al.}, ``A stochastic geometry model for multi-hop highway
  vehicular communication,'' \emph{IEEE Transactions on Wireless
  Communications}, vol.~15, no.~3, pp. 2276--2291, 2015.

\bibitem{andrews2016primer}
J.~G. Andrews \emph{et~al.}, ``A primer on cellular network analysis using
  stochastic geometry,'' \emph{arXiv preprint arXiv:1604.03183}, 2016.

\bibitem{di2015stochastic}
M.~Di~Renzo, ``Stochastic geometry modeling and analysis of multi-tier
  millimeter wave cellular networks,'' \emph{IEEE Transactions on Wireless
  Communications}, vol.~14, no.~9, pp. 5038--5057, 2015.

\bibitem{baccelli1997stochastic}
F.~Baccelli \emph{et~al.}, ``Stochastic geometry and architecture of
  communication networks,'' \emph{Telecommunication Systems}, vol.~7, no.~1,
  pp. 209--227, 1997.

\bibitem{6260478}
F.~Morlot, ``{A population model based on a Poisson line tessellation},'' in
  \emph{{IEEE WiOpt}}, May 2012, pp. 337--342.

\bibitem{chetlur2020stochastic}
V.~V. Chetlur~Ravi, ``Stochastic geometry for vehicular networks,'' Ph.D.
  dissertation, Virginia Tech, 2020.

\bibitem{jeyaraj2020cox}
J.~P. Jeyaraj and M.~Haenggi, ``Cox models for vehicular networks: Sir
  performance and equivalence,'' \emph{IEEE Transactions on Wireless
  Communications}, vol.~20, no.~1, pp. 171--185, 2020.

\bibitem{gloaguen2010analysis}
C.~Gloaguen \emph{et~al.}, ``Analysis of shortest paths and subscriber line
  lengths in telecommunication access networks,'' \emph{Networks and Spatial
  Economics}, vol.~10, no.~1, pp. 15--47, 2010.

\bibitem{choi2017analytical}
C.-S. Choi and F.~Baccelli, ``{An Analytical Framework for Coverage in Cellular
  Networks Leveraging Vehicles},'' \emph{IEEE Trans. Commun.}, vol.~66, no.~10,
  pp. 4950--4964, Oct. 2018.

\bibitem{ghatak2017modeling}
G.~Ghatak \emph{et~al.}, ``Modeling and analysis of hetnets with mm-wave
  multi-rat small cells deployed along roads,'' in \emph{GLOBECOM 2017-2017
  IEEE Global Communications Conference}.\hskip 1em plus 0.5em minus
  0.4em\relax IEEE, 2017, pp. 1--7.

\bibitem{ghatak2019small}
------, ``Small cell deployment along roads: Coverage analysis and slice-aware
  rat selection,'' \emph{IEEE Transactions on Communications}, vol.~67, no.~8,
  pp. 5875--5891, 2019.

\end{thebibliography}
\bibliographystyle{IEEEtran}
\end{document}